\documentstyle[11pt,psfig]{article}
\begin{document}
\baselineskip 0.6cm
\newcommand{\gsim}{ \mathop{}_{\textstyle \sim}^{\textstyle >} }
\newcommand{\lsim}{ \mathop{}_{\textstyle \sim}^{\textstyle <} }
\newcommand{\vev}[1]{ \left\langle {#1} \right\rangle }
\newcommand{\bra}[1]{ \langle {#1} | }
\newcommand{\ket}[1]{ | {#1} \rangle }
\newcommand{\ev}{ {\rm eV} }
\newcommand{\kev}{ {\rm keV} }
\newcommand{\mev}{ {\rm MeV} }
\newcommand{\gev}{ {\rm GeV} }
\newcommand{\tev}{ {\rm TeV} }
\newcommand{\mpl}{$M_{Pl}$}
\newcommand{\mw}{$M_{W}$}

\begin{titlepage}

\begin{flushright}
UW/PT-01/14 \\
\end{flushright}

\vskip 0.4cm

\begin{center}
{\Large \bf  Ineffective Supersymmetry:
\\ Electroweak Symmetry Breaking from Extra Dimensions}

\vskip 1.0cm

\def\thefootnote{\fnsymbol{footnote}}
{\bf 
Neal Weiner$^{\dagger}$
}

\vskip 0.5cm

$^\dagger$ {\it Department of Physics, University of Washington, 
                 Seattle, WA 98195}\\

\vskip 1.0cm

\abstract{Recently, a mechanism for electroweak symmetry breaking 
(EWSB) was discussed \cite{Arkani-Hamed:2001mi}, in which the scale of EWSB is set by 
the scale of an additional dimension $R\sim \tev^{-1}$. The mechanism 
involves supersymmetry, but broken in such a fashion 
that high (four-dimensional) momentum loops are cut off by the finite 
size of the radius. In a Kaluza-Klein decomposition, a hard cutoff 
seems to give a strong cutoff dependence, while summing the entire 
tower is not only cutoff insensitive, but actually finite. Such 
behavior is easily understood in a formulation that respects 
five-dimensional locality. Finally, we note that certain models of 
this type naturally give operators which can ``fake'' the presence of 
a light Higgs in precision electroweak observables.}

\end{center}
\end{titlepage}

\section{Introduction}
A central question in particle physics is this: what is the origin of the 
scale of electroweak symmetry breaking, and what stabilizes it against 
radiative corrections? In the past two decades, a tremendous amount 
of effort has gone into answering this question, and exploring the 
implications of possible answers. This ``hierarchy problem'' is summed 
up simply: why is $M_{W}\ll M_{Pl}$?

Numerous proposals have been made to stabilize the weak scale against 
radiative corrections, two notable examples being technicolor 
\cite{Chivukula:2000mb} and weak-scale supersymmetry \cite{Martin:1997ns}. In technicolor, 
divergences associated with the Higgs sector are controlled because 
the Higgs is not a fundamental field in the theory. In supersymmetric 
theories, above the supersymmetry breaking scale, bosonic and 
fermionic loops cancel, leading to relative insensitivity to the 
cutoff of the theory.

More recently, 
the question of the hierarchy problem has been recast in theories 
with large extra dimensions \cite{Arkani-Hamed:1998rs,Antoniadis:1998ig}. It was noted that if there 
are $n$ extra dimensions, the effective, four-dimensional Planck 
scale is related to the higher dimensional Planck scale $M_{*}$, by 
the equation
\begin{equation}
	M_{Pl}^{2}=M_{*}^{2+n} V,
\end{equation}
where $V$ is the volume of the additional $n$ dimensions. If this 
volume is large, $M_{*}$ can be brought down nearly to the weak 
scale. If this is the case, loop divergences of the standard model are 
cut off at the $\tev$ scale. While this is not a solution to the 
hierarchy problem, it recasts it as the question of the origin of the 
large volume of the extra dimensions.

While these proposals offer interesting frameworks within which to 
work, none of them is complete without an understanding of the origin 
of electroweak symmetry breaking. For instance, supersymmetry with 
gravity mediated supersymmetry breaking \cite{Hall:1983iz}
offers an attractive 
understanding of symmetry breaking. If we assume 
general soft breaking masses 
defined at $\mu=$\mpl, 
top/stop loops drive the Higgs soft mass-squared negative in the RG evolution,
triggering EWSB at roughly the scale of supersymmetry breaking.

In contrast, there is no obvious explanation of EWSB in theories with 
large extra dimensions. Various limits require the higher dimensional 
Planck scale to be relatively large compared to \mw, with $M_{*}\sim 
10 \tev$. Without a theory of electroweak symmetry breaking, it is 
impossible even to begin to discuss whether this scale is unnaturally 
high, or, equivalently, to understand when limits on $M_{*}$ become meaningful.

In this paper, we will employ a new mechanism for electroweak symmetry 
breaking appropriate for large extra dimension (LED) theories. 
It will require the presence of 
an additional dimension of size $R^{-1}\sim \tev$, and supersymmetry 
in the full five-dimensional theory. It will be different from 
previous models with $\tev$ sized extra dimensions 
\cite{Antoniadis:1990ew,Antoniadis:1993fh,Benakli:1996ut,Antoniadis:1998zg,
Pomarol:1998sd,Antoniadis:1998sd,Delgado:1998qr}, in that both
gauge fields and 
matter fields - $Q,U,D,L$ and $E$ - will all propagate in five 
dimensions. Yukawas will be present on one brane, the ``Yukawa brane'', 
while supersymmetry breaking will arise either localized at some distance 
from the Yukawa brane (the SUSY-breaking brane), or through global 
properties of the five dimensional space, as in the Scherk-Schwarz 
mechanism \cite{Scherk:1979ta}. Because high momentum loops will be unable to 
sense both supersymmetry breaking and the Yukawas simultaneously, 
contributions to the Higgs mass will be insensitive to the cutoff of 
the theory. In fact, in contrast with ordinary four-dimensional 
supersymmetry theories, in these theories the Higgs soft mass
will be {\em finite} in the UV!

\subsection{Supersymmetry {\em and} Extra Dimenions?}
At first glance, it may appear that using both supersymmetry and large 
extra dimensions to address the hierarchy problem is somehow 
redundant. After all, isn't supersymmetry on its own a solution to the 
hierarchy problem?

The answer, of course, is that {\em weakly broken} supersymmetry is a 
solution to the hierarchy problem, rather than supersymmetry itself. 
Given a supersymmetric theory, one needs an additional sector to 
generate an exponentially small scale for supersymmetry breaking. 
Only then does one have a complete solution to the hierarchy problem.

Here, we would argue that we have merely replaced sector generating 
the small scale of supersymmetry breaking with a sector which 
generates an exponentially large volume for the extra dimensions, or, 
equivalently, the exponentially small scale $V^{-1/n}$. Then 
supersymmetry can be broken at order one.

Moreover, part of our motivation for considering extra dimensions 
is the presence in superstring theory. Since supersymmetry appears in 
these theories, it is worthwhile to consider the effects of 
supersymmetry in LED theories.

\section{Features of Five-Dimensional Theories}
Let us briefly review the standard formalism for approaching five 
dimensional theories.

To begin with, let us consider the Klein-Gordon equation for a scalar 
field in five dimensions:
\begin{equation}
	(\partial_{t}^{2}-\partial_{\vec x}^{2}-\partial_{y}^{2}) 
	\phi(t,{\vec x},y) = 0,
\end{equation}
where $y$ is the coordinate of a compact fifth dimension, with radius 
$R$. Because the fifth dimension is compact, the values for the 
momentum in the fifth direction can only take on discrete values, $0, 
R^{-1}, 2 R^{-1},$ etc. When we Fourier decompose $\phi(t,{\vec 
x},y)$, the Klein-Gordon equation becomes
\begin{equation}
	(\partial_{t}^{2}-\partial_{\vec x}^{2}+k^{2} R^{-2}) 
	\phi_{k}(t,{\vec x}) = 0.
\end{equation}
Thus each Fourier mode (or, equivalently, each Kaluza-Klein mode), 
acts as a four-dimensional field with mass $k/R$. The $y$ dependence 
has been replaced by the values of an infinite tower of fields. 
Although we have used scalars here as an example, the same is true for 
higher spin particles.

If additional dimensions are present, we expect to see Kaluza-Klein 
towers for the fields which propagate in them. For standard-model 
fields, the absence of four-fermion operators, and corrections to 
precision electroweak observables generally constrain these dimensions 
to be somewhat small $R^{-1} \sim \tev$ \cite{Barbieri:2000vh}. In contrast, 
gravity can propagate in dimensions as large as $R \sim 10^{-5}m$ 
\cite{Hoyle:2000cv}.

\subsection{Supersymmetry in Five Dimensions}
If we wish to discuss five dimensional supersymmetric theories, we 
need to understand the differences between four- and five-dimensional 
supersymmetry. Supersymmetry has a fermionic generator, and 
the minimal fermionic representation of the five-dimensional Lorentz 
group is a four-component spinor. This spinor decomposes under the 
four-dimensional Lorentz group as two two-component Weyl spinors. 
Thus, we see that five-dimensional $N=1$ supersymmetry is equivalent 
to four-dimensional $N=2$ supersymmetry.

In $N=2$ supersymmetry, the minimal representations of the 
supersymmetry algebra have more components than in $N=1$ 
\cite{Arkani-Hamed:2001tb}. For instance, chiral superfields become part of 
hypermultiplets which contain two chiral superfields. Moreover, from a 
four-dimensional point of view, these superfields are vectorlike under 
gauge symmetries. That is, given a hypermultiplet $(\Phi, \Phi^{c})$, 
if $\Phi$ transforms under $SU(n)$ as an ${\bf N}$, then $\Phi^{c}$ 
transforms as an ${\bf \overline N}$. 

Similarly, gauge field multiplets no longer consist merely of a vector 
and a fermion. They now contain an additional chiral superfield which 
transforms as an adjoint under the gauge group.

In addition to enlarging the field content of the theory, $N=2$ 
supersymmetry strongly constrains the form of the interactions we can  
write. In particular, we cannot write trilinear couplings, such as 
Yukawas. That is, terms like
\begin{equation}
	\int d^{4}x dy d^{2}\theta \> f_{t} Q U H
	\label{eq:bulkyukawa}
\end{equation}
are forbidden. 

All of these issues seem problematic phenomenologically. For one 
thing, we know that matter is chiral, not vectorlike under the gauge 
groups. Where are these $Q^{c},U^{c},D^{c},L^{c}$ and $E^{c}$ states? 
We do not see massless chiral adjoint fields, and the masses of 
fermions seem to require the presence of trilinear interactions. How 
can we resolve these problems?

Let us separate these into separate issues: missing states and needed 
couplings. The missing states are typically removed by use of an 
``orbifold projection''. Simply put it is this: our Kaluza-Klein 
(Fourier) decomposition can be written in terms of even functions 
(cosines) and odd functions (sines). It is a consistent projection to 
require that both the conjugate states and chiral adjoint fields 
satisfy odd boundary conditions (i.e., only keep the sines), while the 
unconjugated fields and vector supersuperfields satisfy even boundary 
conditions (i.e., only keep the cosines). Then there are no massless 
chiral adjoints, and there are no massless partners of the chiral 
matter fields.

Under this orbifold projection, two points on the circle are special, 
$y=0$ and $y=\pi R$. These ``fixed points'' are mapped to themselves 
under the transformation $y\rightarrow -y$. We can naturally place 
branes at these points, and the interactions on these branes need 
only preserve four-dimensional Lorentz invariance, and hence, only 
$N=1$ supersymmetry. That is, while eq. \ref{eq:bulkyukawa} is 
forbidden, a term
\begin{equation}
	\int d^{4}x dy d^{2}\theta \delta(y) f_t Q U H
\end{equation}
is allowed.

\section{Breaking Supersymmetry}
We can now explore the interesting possibility: what if supersymmetry 
is broken ``away'' from the Yukawa brane? This can take different 
forms, and let us consider two distinct possibilities. First, perhaps 
supersymmetry is broken by an F-component expectation value of some 
chiral superfield $X$ localized on the brane at $y=\pi R$. The second 
possibility is that we use the Scherk-Schwarz mechanism, and require 
different boundary conditions for particles and their superpartners.

We will focus our attention on the localized supersymmetry breaking 
scenario, and return to the case of Scherk-Schwarz models shortly.

Suppose that  $F_{X}\sim M_{*}^{2}$ is localized at $y=\pi R$. Then 
interactions (setting $M_{*}=1$)
\begin{equation}
	\int d^{4}x dy d^{4}\theta \> 
	c_{X} X^{\dagger} X Q^{\dagger} Q \delta(y-\pi R)
	\label{eq:softmass}
\end{equation}
(and similar terms for $U,D,L,E$)
will generate soft masses for superpartners. How large are these 
masses? If we use the naive Kaluza Klein decomposition, we get mass 
terms
\begin{equation}
	\int d^{4}x c_{X} F_{X}^{2} R^{-1} \tilde q^{*}_{i} \tilde q_{j}.
\end{equation}
Focusing on the diagonal ($i=j$) entries of this, we might expect 
that all superpartners have masses $\sim F_{X} \sqrt{c_{X} R^{-1}}$ 
(again in units 
with $M_{*}=1$). However, this assumes that such an operator is a small 
perturbation on the system and it is not. To find the mass spectrum, 
we should instead solve the differential equation
\begin{equation}
	-\partial_{y}^{2} g(y) + m^{2} g(y) + c_{X} F_{X}^{2}
	\delta(y-\pi R) g(y) =0.
\end{equation}
Solving this for even $g(y)$, we have solutions
\begin{equation}
	g(y) = C \cos(m y)
\end{equation}
where $m$ is a solution of the equation 
\begin{equation}
	\tan (m \pi R) = \frac{c_{X} F_{X}^{2}}{2 m}.
\end{equation}
For $R \gg 1$ in fundamental units, this has solutions
\begin{equation}
	m \simeq (n+\frac{1}{2})R^{-1}.
\end{equation}
That is, we have shifted the Kaluza-Klein spectrum of our 
superpartners up one-half unit in $R^{-1}$ \footnote{The analysis for 
gauginos is somewhat different, and is carried out in \cite{Arkani-Hamed:2001mi}.}. 
In fact, at this
(leading) order in $R^{-1}$, the spectrum is independent of the size 
$c_{X}$ so long as it is order one! 
Flavor changing effects arising from superpartner loops are naturally 
suppressed by this high degree of degeneracy. 
In fact, it is easy to understand this behavior: the 
large mass term in eq. \ref{eq:softmass} attempts to give large 
masses to the superpartners. By avoiding the brane at $y=\pi R$ (that 
is, by developing a node at $y=\pi R$), the 
superpartners pick up masses $m \sim R^{-1}$ from the kinetic terms, 
but no mass from the brane. 
So long as $R^{-1} < F_{X} \sqrt{c_{X} R^{-1}}$, or, 
equivalently, $R^{-1} < F_{X}^{2} c_{x}$, the masses of the lightest 
modes are smaller with an approximate node at the brane at $y=\pi R$. 

\begin{figure}
\begin{center} 
\centerline{\psfig{file=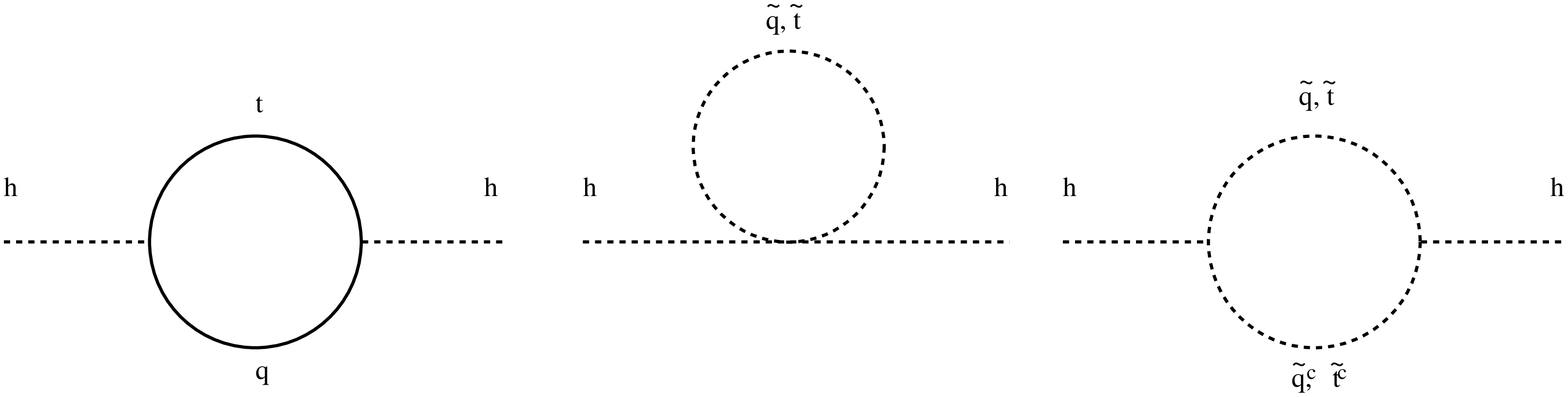,width=0.8\textwidth}}
\caption{One-loop diagrams contributing to the Higgs-boson mass.}
\label{fig:Fig_RC-Higgs}
\end{center}
\end{figure}

Now that we have found the spectrum of our superpartners, we can 
calculate loops corrections to the soft mass of the Higgs field, 
which we show in figure \ref{fig:Fig_RC-Higgs}. Note the presence of 
the new diagram involving the conjugate fields. This diagram is 
necessary to ensure a cancellation in the supersymmetric limit (and is 
easily derived from the boundary Yukawa with a bulk superpotential 
term $\int dy Q^{c} \partial_{y} Q$ \cite{Arkani-Hamed:2001tb,Arkani-Hamed:1999pv}).

We now have the spectrum of the theory, but to calculate the loop 
diagrams, we must properly normalize the modes.
The wavefunctions $g_k(y)$ for the KK modes are normalized such that 
$\int_0^{\pi R} [g_k(y)]^2 dy = \pi R$. 
If $\eta^i_k$ is the value of the $k$-th Kaluza Klein mode of
the field $i$ on the Yukawa brane, we have 
\begin{eqnarray}
  && \eta^{\psi}_k = (\frac{1}{\sqrt{2}})^{\delta_{k,0}}, \\
  && \eta^{\phi}_k = 1, \\
  && \eta^{F}_k = \left\{ \begin{array}{ll}
    (\frac{1}{\sqrt{2}})^{\delta_{k,0}} 
    \qquad & {\rm for} \quad r^{F}=0,\\ 
    1 \qquad & {\rm for} \quad r^{F}=\frac{1}{2},\\ 
  \end{array} \right. \label{eq:ee3}
\end{eqnarray}
where $k = 0,1,2,$ etc.

With these normalizations, we can calculate the loop contributions to 
the Higgs soft mass.
Using the variable $x = p_E R$, one can show
\begin{eqnarray}
-i\, m_{\phi_H}^2 &=& \frac{i N_c f_t^2 \epsilon^2}{R^2} 
    \int \frac{d^4 x}{(2\pi)^4} x^2 
\nonumber\\
&&  \times
    \sum_{k,l=0}^{\infty}
    \Biggl[ \frac{(\eta^{\psi}_k)^2 (\eta^{\psi}_l)^2}
      {(x^2+k^2)(x^2+l^2)} 
    - \frac{(\eta^{\phi}_k)^2 (\eta^{F}_l)^2}
      {(x^2+(k+r^{\phi})^2)(x^2+(l+r^{F})^2)} \Biggr].
\label{eq:sum-mass}
\end{eqnarray}

One might worry that each mode would 
contribute to the Higgs soft mass, and thus give a severe cutoff 
dependence. However, this is not the case: we can in fact sum the 
entire tower and achieve the {\em finite} result
\begin{equation}
	m_{H_{u}}^{2} = - \frac{3 \zeta(3)}{8 \pi^{4}}\frac{N_{c} 
	y_{t}^{2}}{R^{2}}.
\end{equation}
This result seems remarkable! It is especially so when we note that 
the lightest stops do not have the couplings appropriate for an 
ordinary four-dimensional SUSY theory. That is, if one only saw the 
lightest sfermions, their couplings to the Higgs differ from those 
expected from four-dimensional $N=1$ supersymmetry by a factor 
$\sqrt{2}$. It is tempting to refer to this scenario as 
``ineffective supersymmetry.'' While supersymmetry is protecting the 
Higgs mass, at no energy do we ever encounter a four-dimensional 
effective supersymmetric theory.

Here we have summed the whole tower. If we had employed a hard cutoff in 
\ref{eq:sum-mass}, we would have found a severe cutoff 
dependence, in accord with naive expectations. How can we be certain 
that this technique (summing to infinity) is the right one?

Within the Kaluza-Klein formalism, it is somewhat difficult to see 
that this is correct. In fact, there has been a great deal of 
discussion recently
\cite{Ghilencea:2001ug,Delgado:2001ex,Contino:2001gz,Nomura:2001ec}, 
as to whether such a 
summation amounts to a hidden fine tuning. What can we say?

A hard cutoff is clearly 
inappropriate. Since the physical features of the theory should 
determine its UV properties, any regulator we use must preserve 
certain features, namely, supersymmetry, five-dimensional Lorentz 
invariance, and, importantly, locality. To understand the finite 
nature of the result, we should formulate the calculation in which 
locality is manifest. Indeed, in the next section, we shall see from a 
five-dimensional perspective that the result {\em must} be finite.

\section{Five dimensional interpretation}
It is easiest to see that the result must be finite in a formalism 
which makes the fifth dimension explicit \cite{Arkani-Hamed:2001mi}. Thus, we shall
work in mixed 
position-momentum space \cite{Arkani-Hamed:1999za}. Since 
contributions to the Higgs soft mass rely both on Yukawas localized 
at $y=0$ {\em and} supersymmetry breaking at $y=\pi R$, high momentum 
loops cannot simultaneously ``see'' both elements, causing an 
exponetial damping at high $p_{E}$.

One can calculate the scalar, fermion and F-component propogators 
easily \cite{Arkani-Hamed:2001mi}. In the large susy breaking limit, they are
\begin{equation}
  \tilde G_{\phi}(k_{4},y) = \frac{1}{2 k_{4}} 
  \frac{1}{\sinh[k_4 \pi R]} \left\{ \cosh[k_4 (\pi R - y)] - 
  \frac{m \cosh[k_4 y]}{2 k_4 \sinh[k_4 \pi R] + m \cosh[k_4 \pi R]}
  \right\},
\label{eq:genprop}
\end{equation}
for $(y \in [0, \pi R])$.
Again, we can take the large susy breaking 
limit, $m \rightarrow \infty$ and then 
the propagator becomes
\begin{equation}
  \tilde G_{\phi}(k_{4},y) = \frac{1}{2 k_{4}} 
  \frac{1}{\sinh[k_4 \pi R]} \left\{ \cosh[k_4 (\pi R - y)] - 
  \frac{\cosh[k_4 y]}{\cosh[k_4 \pi R]} \right\},
\label{eq:opaprop}
\end{equation}
\begin{equation}
  \tilde G_{\psi}(k_{4},y) = \frac{\not k_{4}}{2 k_{4}} 
  \frac{\cosh[k_{4}(\pi R- y)]}{\sinh[k_{4}\pi R]},
\label{eq:prop-fermion}
\end{equation}
and
\begin{equation}
  \tilde G_{F}(k_{4},y) = \frac{k_{4}}{2}e^{- k_{4}|y|}.
\end{equation}
\begin{figure}
  \centerline{
  \psfig{file=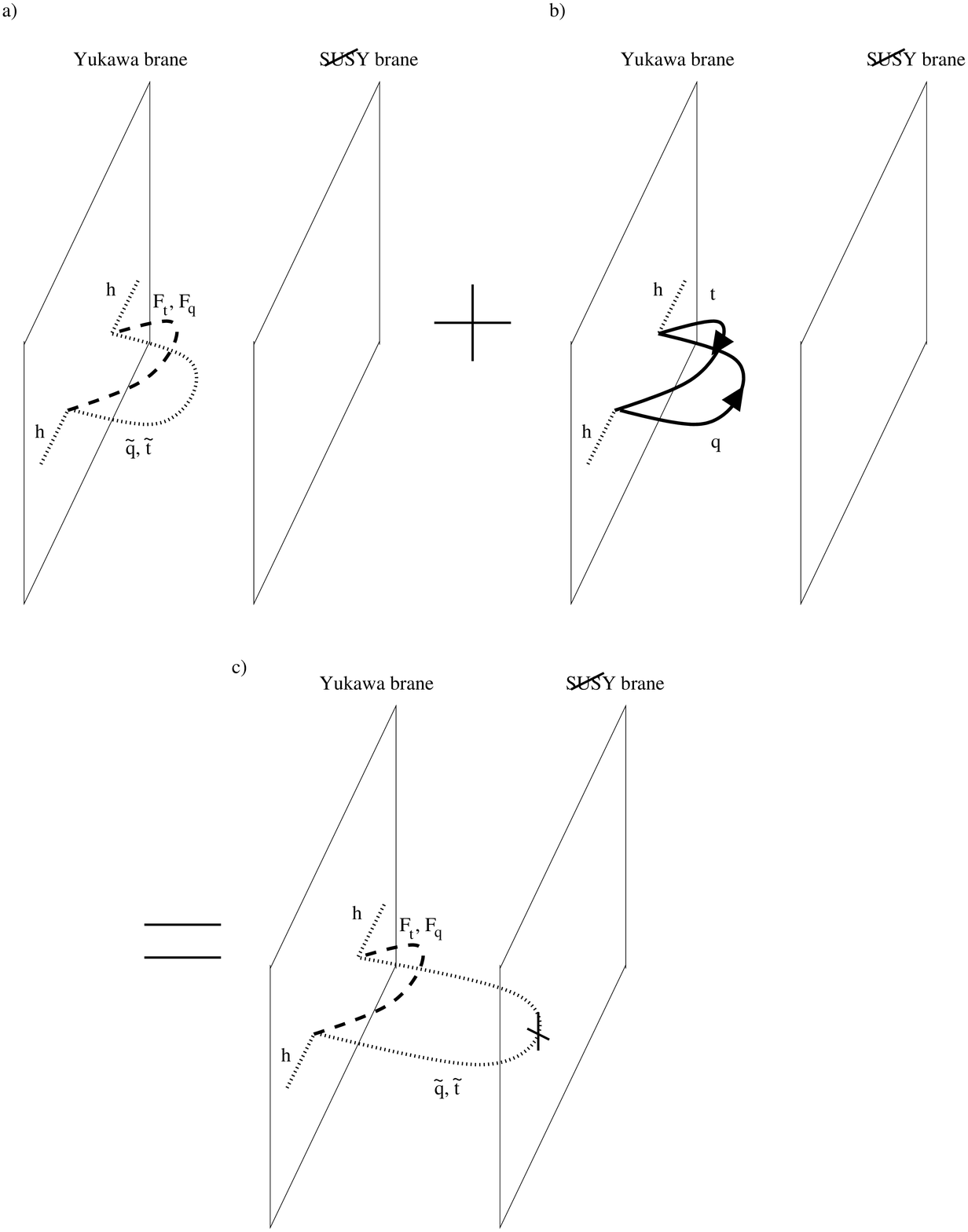,width=0.8\textwidth}}
\caption{Bosonic (a) and fermionic (b) contributions to the 
Higgs soft mass, when added leave only the reflected piece (c).}
\label{fig:5ddiags}
\end{figure}
The bosonic amplitude from Fig.~\ref{fig:5ddiags}(a) is 
given by
\begin{equation}
  m^{2}_{\rm boson} = 2 N_c \int \frac{d^{4}k_{4}}{(2 \pi)^{4}}
  (2 \pi R y_{t})^{2} \frac{\tanh[k_{4}\pi R]}{2 k_{4}}
  \frac{k_{4} \coth[k_{4} \pi R]}{2}, 
\end{equation}
and the fermionic one from Fig.~\ref{fig:5ddiags}(b) by
\begin{equation}
  m^{2}_{\rm fermion} = - N_c \int \frac{d^{4}k_{4}}{(2 \pi)^{4}}
  (2 \pi R y_{t})^{2} {\rm Tr}
  \left[\frac{\not k_{4} \coth[k_{4} \pi R]}{2 k_{4}}
  \frac{(1-\gamma_{5})}{2}
  \frac{\not k_{4} \coth[k_{4} \pi R]}{2 k_{4}}
  \frac{(1+\gamma_{5})}{2}\right].
\end{equation}
These are combined to give a total amplitude
\begin{equation}
  m^{2}_{\rm tot} = 2 N_c \int \frac{d^{4}k_{4}}{(2 \pi)^{4}}
  (2 \pi R y_{t})^{2} \frac{\coth[k_{4} \pi R]}{4} 
  \left( \tanh[k_{4} \pi R] - \coth[k_{4} \pi R] \right).
\end{equation}
We can rewrite this as
\begin{equation}
  m^{2}_{\rm tot} = - \frac{N_c\, y_{t}^{2}}{4 R^{2}}
  \int_0^{\infty}dx \frac{x^{3}}{\sinh^{2}[\pi x]} = 
  - \frac{3\, \zeta(3)}{8 \pi^{4}} \frac{N_c\, y_{t}^{2}}{R^{2}},
  \label{eq:result}
\end{equation}
precisely what we found before.

The finiteness is now easily understood: since loops originate on the 
Yukawa brane, any loops in the vicinity of the Yukawa brane are 
supersymmetric and cancel. Loops which probe the bulk, and reflect 
off the supersymmetry breaking brane will not cancel. However, these 
must vanish in the UV, leaving a finite result. We see this 
schematically in figure \ref{fig:5ddiags}.

In fact, with this understanding, we need not have calculated the 
fermionic loop at all! One can simply calculate the 
scalar/F-component diagram, and subtract off the value in the 
$m\rightarrow 0$ limit, which yields eq. \ref{eq:result} again. This 
verifies our understanding of the finite result.

\subsection{Scherk-Schwarz Theories}
We have seen that supersymmetry breaking, localized away from the 
Yukawa brane can lead to UV-finite contributions. However, another 
alternative exists, utilizing the {\em global} properties of the 
five-dimensional space, via the Scherk-Schwarz mechanism \cite{Scherk:1979ta}. 

Models can be constructed which give different boundary conditions to 
standard model particles and their superpartners 
(for instance, \cite{Arkani-Hamed:2001mi,Pomarol:1998sd,Barbieri:2000vh}). 
We will restrict our brief discussion to the model of \cite{Arkani-Hamed:2001mi}
in which $R$-parity assignments classify which modes are even 
and which are odd.

Under $R$ parity, various superfields transform as
\begin{eqnarray}
\begin{array}{ll}
  X(x, y, \theta)      \rightarrow  -X(x, y, -\theta), \qquad &
  X^c(x, y, \theta)    \rightarrow  -X^c(x, y, -\theta),
\nonumber\\
  H(x, y, \theta)      \rightarrow   H(x, y, -\theta), \qquad &
  H^c(x, y, \theta)    \rightarrow   H^c(x, y, -\theta),
\\
  V(x, y, \theta)      \rightarrow   V(x, y, -\theta), \qquad &
  \Sigma(x, y, \theta) \rightarrow   \Sigma(x, y, -\theta),
\nonumber
\end{array}
\end{eqnarray}
where $X$ and $H$ represent $Q, U, D, L, E$ and $H_u, H_d$, respectively. 
It should be understood that the Higgs field need not propogate in the 
bulk, but we have listed its transformation properties in the event 
that it does.

Fermionic tops then have the ordinary Kaluza-Klein tower, $m = 0, 
R^{-1}, 2 R^{-1},$ etc., and fermionic top conjugates remain unchanged from 
the intial orbifold, $m = R^{-1}, 2 R^{-1}, 3 R^{-1},$ etc. In 
contrast, stops and conjugate stops are shifted one unit up and down, 
respectively, $m= R^{-1}/2, 3R^{-1}/2, 5 R^{-1}/2,$ etc.

We can again calculate the contribution to the Higgs soft mass, and 
summing the entire tower yields the finite result
\begin{equation}
  m_{\phi_{H_u}}^2 = - {21\, \zeta(3) \over 32 \pi^4} N_c\, y_t^2 M_c^2.
\label{eq:HG-2}
\end{equation}

\subsection{Five dimensional interpretation}
Again, we have summed the entire tower of states. Is this the correct 
thing to do? In the previous example, 
we could understand the finite behavior because of the 
locality of the supersymmetry breaking. What is the equivalent five 
dimensional understanding here?

The key point is that a Scherk-Schwarz breaking of supersymmetry is a 
{\em global} feature of the space, not a local one. Thus, short 
distance (i.e., high four-momentum) loops will only probe the local 
features of the spacetime, meaning those short distance loops should 
retain supersymmetric cancellations.

In mixed position-momentum space, this arises because of the different 
properties of particles and their superpartners as we continue around 
the extra dimension. That is, under $y \rightarrow y+ 2 \pi R$, the 
squarks will pick up a minus sign in their propagator.
\begin{equation}
  \tilde G_{\phi}(k_{4},y) 
  = \sum_{n=-\infty}^{\infty} \frac{1}{2 k_{4}} (-1)^{n} 
    e^{-k_{4} |y - 2 \pi n R|}.
	\label{eq:ssscalarprop}
\end{equation}
At the same time, the fermion picks up a plus sign when $y \rightarrow y+ 
2 \pi R$. 
\begin{equation}
  \tilde G_{\psi}(k_{4},y) 
  = \sum_{n=-\infty}^{\infty} \frac{\not k_{4}}{2 k_{4}}
    e^{-k_{4} |y - 2 \pi n R|}.
\label{eq:ssfermionprop}
\end{equation}

Note that the supersymmetry breaking occurs in these propagators only 
when the scalar and fermion both propagate {\em completely around the 
extra dimension} (i.e., $n \ne 0$). 
High four-momentum loops will exponentially damp 
out in this region, leaving only the supersymmetrically cancelling 
pieces of the propagators. A full calculation has been given in 
\cite{Arkani-Hamed:2001mi}. From this formulation, it is clear that the final 
result must be finite in the UV.

\section{Phenomenology}
We shall not delve deeply into the phenomenology of these models, but 
only comment briefly on their properties. Most notably, the squarks 
and sleptons are all degenerate up to corrections from 
electroweak symmetry breaking. The gauginos are degenerate with the 
sfermions in the 
Scherk-Schwarz case, and can be degenerate, but are not necessarily so,
in the localized supersymmetry breaking case. 

Perhaps the most exciting change in phenomenology from standard SUSY 
theories is the change in the Higgs sector. If the Higgsino is the 
NLSP (with localized supersymmetry breaking, the gravitino is the 
LSP), and if the gluino is light enough to be produced at hadron 
colliders, it will decay $\tilde g \rightarrow \overline q q \tilde 
h$, followed by $\tilde h \rightarrow h \tilde G$. If the gluino is 
produced in large enough quantities, this could be the dominant 
mechanism for Higgs production.

So far we have not given a mechanism for generating $\mu$. There are 
many possibilities. An attractive example is to utilize a next 
to minimal sector as in the NMSSM with a superpotential
\begin{equation}
  W_{\rm Higgs} = \lambda_H S H_{u}H_{d} + \lambda_B S B \bar{B}
  +{\lambda_S \over 3}S^3. 
\end{equation}
where $S$ is a brane field and $B, \overline B$ are bulk fields. 
Just as the Higgs soft mass is driven negative, so is the $S$ soft 
mass, giving it a vev and generating a $\mu$ term. Ordinarily, one 
must be concerned that the trilinear couplings will run strong before 
the Planck scale. Here, the cutoff is only a few $\tev$, so we have 
great freedom in choosing the $\lambda$'s. Because of this, it is 
quite easy to have a Higgs much heavier than the $O(130 \gev)$ 
supersymmetric Higgs.

Precision electroweak studies have shown that a heavy Higgs is 
disfavored. However, in \cite{Hall:1999fe} it was shown that the 
inclusion of two nonrenormalizable operators, 
$(H^{\dagger}\tau^{a}H)W_{\mu \nu}^{a}B_{\mu \nu}$ and 
$|H^{\dagger}D_{\mu}H|^{2}$,
could ``fake'' the existence of a light Higgs boson, at least with 
regard to precision electroweak studies. In \cite{Barbieri:1999tm} it 
was pointed out that if one allowed all possible operators 
uncontrolled by flavor symmetries at the same level (specifically 
$(H^{\dagger}D_{\mu}\tau^{a}H)(\overline X 
\gamma_{\mu}\tau^{a}X)_{X\rightarrow L,Q}$,
$(H^{\dagger}D_{\mu}H)(\overline X \gamma_{\mu}X)_{X\rightarrow Q,U,D,LE}$ 
, and $(\overline L \gamma_{\mu} \tau^{a} L)^{2}$), then
to fake a light 
Higgs was incompatible with other precision studies. Put another way: 
any model that seeks to explain the precision studies of $m_H$ with these 
nonrenomralizable operators must explain why the operators of 
\cite{Hall:1999fe} are 
anomalously large when compared with a host of other operators.

However, we note that all the operators which need be suppressed 
involve fields which live in the bulk. Thus, these operators are all 
naturally suppressed by a volume factor $1/2\pi R$ relative to the 
operators discussed by Hall and Kolda. The operator involving only 
the gauge bosons and the Higgs 
does not receive volume suppression, which simply 
goes into defining the four-dimensional gauge couplings from the 
five-dimensional gauge couplings. Thus, here we have precisely an 
example of a model in which the ``bad'' operators are automatically 
suppressed, while the ``good'' operators remain large. It is then 
quite natural in models where matter fields propagate in the bulk but 
the Higgs lies on a brane to give the appearance in precision tests 
that the Higgs is lighter than it is.

\section{Conclusions}
In theories in which the cutoff scale is $O(\tev)$, we cannot truly 
say we have understood the hierarchy problem without understanding the 
origin of electroweak symmetry breaking. Here we have noted that with 
additional dimension of size $R^{-1}\sim \tev$ and five-dimensional 
supersymmetry, there can be a natural understanding of electroweak 
symmetry breaking. In this mechanism, the Higgs mass is calculably 
finite, and a loop factor below the masses of the superpartners, 
ameliorating fine-tuning issues.

The loops contributing to the Higgs soft mass have an extreme UV 
softness. The five-dimensional understanding of these models gives us an 
intuitive understanding - and calculationally tractable method - of 
understanding the origin of this behavior.

The phenomenology of these models is rich, containing heavy, degenerate 
superpartners. The Higgs sector can be radically changed. Not only are 
the theoretical controls on the light Higgs boson mass lifted, so, 
too, are those constraints from precision electroweak measurements. 
This makes the direct search for the Higgs boson in regions 
$M_{H}>140 \gev$ especially interesting.

\setcounter{footnote}{0}


\section*{Acknowledgements}

I would like to thank the organizers of the 2001 Rencontres de Moriond: 
Electroweak Interactions and Unified Theories. I am indebted to Nima 
Arkani-Hamed, Lawrence Hall, Yasunori Nomura and David Smith for 
conversations and collaboration on the work on which this is based. 
This work was partially supported by the 
DOE under contract DE-FG03-96-ER40956.
\bibliography{moriond}

\providecommand{\href}[2]{#2}\begingroup\raggedright\begin{thebibliography}{10}

\bibitem{Arkani-Hamed:2001mi}
N.~Arkani-Hamed, L.~Hall, Y.~Nomura, D.~Smith, and N.~Weiner, {\it Finite
  radiative electroweak symmetry breaking from the bulk},
  \href{http://xxx.lanl.gov/abs/hep-ph/0102090}{{\tt hep-ph/0102090}}.

\bibitem{Chivukula:2000mb}
R.~S. Chivukula, {\it Lectures on technicolor and compositeness},
  \href{http://xxx.lanl.gov/abs/hep-ph/0011264}{{\tt hep-ph/0011264}}.

\bibitem{Martin:1997ns}
S.~P. Martin, {\it A supersymmetry primer},
  \href{http://xxx.lanl.gov/abs/hep-ph/9709356}{{\tt hep-ph/9709356}}.

\bibitem{Arkani-Hamed:1998rs}
N.~Arkani-Hamed, S.~Dimopoulos, and G.~Dvali, {\it The hierarchy problem and
  new dimensions at a millimeter},  {\em Phys. Lett.} {\bf B429} (1998)
  263--272, [\href{http://xxx.lanl.gov/abs/hep-ph/9803315}{{\tt
  hep-ph/9803315}}].

\bibitem{Antoniadis:1998ig}
I.~Antoniadis, N.~Arkani-Hamed, S.~Dimopoulos, and G.~Dvali, {\it New
  dimensions at a millimeter to a fermi and superstrings at a tev},  {\em Phys.
  Lett.} {\bf B436} (1998) 257--263,
  [\href{http://xxx.lanl.gov/abs/hep-ph/9804398}{{\tt hep-ph/9804398}}].

\bibitem{Hall:1983iz}
L.~Hall, J.~Lykken, and S.~Weinberg, {\it Supergravity as the messenger of
  supersymmetry breaking},  {\em Phys. Rev.} {\bf D27} (1983) 2359--2378.

\bibitem{Antoniadis:1990ew}
I.~Antoniadis, {\it A possible new dimension at a few tev},  {\em Phys. Lett.}
  {\bf B246} (1990) 377--384.

\bibitem{Antoniadis:1993fh}
I.~Antoniadis, C.~Munoz, and M.~Quiros, {\it Dynamical supersymmetry breaking
  with a large internal dimension},  {\em Nucl. Phys.} {\bf B397} (1993)
  515--538, [\href{http://xxx.lanl.gov/abs/hep-ph/9211309}{{\tt
  hep-ph/9211309}}].

\bibitem{Benakli:1996ut}
K.~Benakli, {\it Perturbative supersymmetry breaking in orbifolds with wilson
  line backgrounds},  {\em Phys. Lett.} {\bf B386} (1996) 106--114,
  [\href{http://xxx.lanl.gov/abs/hep-th/9509115}{{\tt hep-th/9509115}}].

\bibitem{Antoniadis:1998zg}
I.~Antoniadis, S.~Dimopoulos, and G.~Dvali, {\it Millimeter range forces in
  superstring theories with weak- scale compactification},  {\em Nucl. Phys.}
  {\bf B516} (1998) 70--82, [\href{http://xxx.lanl.gov/abs/hep-ph/9710204}{{\tt
  hep-ph/9710204}}].

\bibitem{Pomarol:1998sd}
A.~Pomarol and M.~Quiros, {\it The standard model from extra dimensions},  {\em
  Phys. Lett.} {\bf B438} (1998) 255--260,
  [\href{http://xxx.lanl.gov/abs/hep-ph/9806263}{{\tt hep-ph/9806263}}].

\bibitem{Antoniadis:1998sd}
I.~Antoniadis, S.~Dimopoulos, A.~Pomarol, and M.~Quiros, {\it Soft masses in
  theories with supersymmetry breaking by tev-compactification},  {\em Nucl.
  Phys.} {\bf B544} (1999) 503--519,
  [\href{http://xxx.lanl.gov/abs/hep-ph/9810410}{{\tt hep-ph/9810410}}].

\bibitem{Delgado:1998qr}
A.~Delgado, A.~Pomarol, and M.~Quiros, {\it Supersymmetry and electroweak
  breaking from extra dimensions at the tev-scale},  {\em Phys. Rev.} {\bf D60}
  (1999) 095008, [\href{http://xxx.lanl.gov/abs/hep-ph/9812489}{{\tt
  hep-ph/9812489}}].

\bibitem{Scherk:1979ta}
J.~Scherk and J.~H. Schwarz, {\it Spontaneous breaking of supersymmetry through
  dimensional reduction},  {\em Phys. Lett.} {\bf B82} (1979) 60.

\bibitem{Barbieri:2000vh}
R.~Barbieri, L.~J. Hall, and Y.~Nomura, {\it A constrained standard model from
  a compact extra dimension},  {\em Phys. Rev.} {\bf D63} (2001) 105007,
  [\href{http://xxx.lanl.gov/abs/hep-ph/0011311}{{\tt hep-ph/0011311}}].

\bibitem{Hoyle:2000cv}
C.~D. Hoyle {\em et.~al.}, {\it Sub-millimeter tests of the gravitational
  inverse-square law: A search for 'large' extra dimensions},  {\em Phys. Rev.
  Lett.} {\bf 86} (2001) 1418--1421,
  [\href{http://xxx.lanl.gov/abs/hep-ph/0011014}{{\tt hep-ph/0011014}}].

\bibitem{Arkani-Hamed:2001tb}
N.~Arkani-Hamed, T.~Gregoire, and J.~Wacker, {\it Higher dimensional
  supersymmetry in 4d superspace},
  \href{http://xxx.lanl.gov/abs/hep-th/0101233}{{\tt hep-th/0101233}}.

\bibitem{Arkani-Hamed:1999pv}
N.~Arkani-Hamed, L.~Hall, D.~Smith, and N.~Weiner, {\it Exponentially small
  supersymmetry breaking from extra dimensions},  {\em Phys. Rev.} {\bf D63}
  (2001) 056003, [\href{http://xxx.lanl.gov/abs/hep-ph/9911421}{{\tt
  hep-ph/9911421}}].

\bibitem{Ghilencea:2001ug}
D.~M. Ghilencea and H.-P. Nilles, {\it Quadratic divergences in kaluza-klein
  theories},  {\em Phys. Lett.} {\bf B507} (2001) 327--335,
  [\href{http://xxx.lanl.gov/abs/hep-ph/0103151}{{\tt hep-ph/0103151}}].

\bibitem{Delgado:2001ex}
A.~Delgado, G.~von Gersdorff, P.~John, and M.~Quiros, {\it One-loop higgs mass
  finiteness in supersymmetric kaluza- klein theories},
  \href{http://xxx.lanl.gov/abs/hep-ph/0104112}{{\tt hep-ph/0104112}}.

\bibitem{Contino:2001gz}
R.~Contino and L.~Pilo, {\it A note on regularization methods in kaluza-klein
  theories},  \href{http://xxx.lanl.gov/abs/hep-ph/0104130}{{\tt
  hep-ph/0104130}}.

\bibitem{Nomura:2001ec}
Y.~Nomura, {\it Constrained standard model from extra dimension},
  \href{http://xxx.lanl.gov/abs/hep-ph/0105113}{{\tt hep-ph/0105113}}.

\bibitem{Arkani-Hamed:1999za}
N.~Arkani-Hamed, Y.~Grossman, and M.~Schmaltz, {\it Split fermions in extra
  dimensions and exponentially small cross-sections at future colliders},  {\em
  Phys. Rev.} {\bf D61} (2000) 115004,
  [\href{http://xxx.lanl.gov/abs/hep-ph/9909411}{{\tt hep-ph/9909411}}].

\bibitem{Hall:1999fe}
L.~Hall and C.~Kolda, {\it Electroweak symmetry breaking and large extra
  dimensions},  {\em Phys. Lett.} {\bf B459} (1999) 213--223,
  [\href{http://xxx.lanl.gov/abs/hep-ph/9904236}{{\tt hep-ph/9904236}}].

\bibitem{Barbieri:1999tm}
R.~Barbieri and A.~Strumia, {\it What is the limit on the higgs mass?},  {\em
  Phys. Lett.} {\bf B462} (1999) 144--149,
  [\href{http://xxx.lanl.gov/abs/hep-ph/9905281}{{\tt hep-ph/9905281}}].

\end{thebibliography}\endgroup
\bibliographystyle{JHEP}

\end{document}